\documentclass[aps,prd,12pt,nofootinbib,superscriptaddress]{revtex4-2}
\usepackage{bm}
\usepackage{latexsym}
\usepackage{amsfonts}
\usepackage{graphicx}
\usepackage{amsmath}
\usepackage{amssymb}
\usepackage{booktabs}
\usepackage{graphicx}
\usepackage{float}
\usepackage{longtable}
\usepackage{subfigure}
\usepackage{comment}
\usepackage{hyperref}
\usepackage{comment}
\usepackage{color}
\usepackage{multirow}
\usepackage{CJKutf8}
\usepackage{orcidlink}
\usepackage{ulem}
\usepackage{CJK}
\hypersetup{
	colorlinks=true,
	linkcolor=red,
	citecolor=blue
}

\begin{document}
\title{Observational constraints on inflationary decoherence with polynomial attractor model}
\author{Zhongkai Wang}
\email{ztlu@hust.edu.cn}
\affiliation{College of Elementary Education, Changsha Normal University, Changsha, Hunan 410100, China}
\affiliation{School of Physics, Huazhong University of Science and Technology, Wuhan, Hubei 430074, China}
\author{Yungui Gong \orcidlink{0000-0001-5065-2259}}
\email{Corresponding author. gongyungui@nbu.edu.cn}
\affiliation{Institute of Fundamental Physics and Quantum Technology, Department of Physics, School of Physical Science and Technology, Ningbo University, Ningbo, Zhejiang 315211, China}
\affiliation{School of Physics, Huazhong University of Science and Technology, Wuhan, Hubei 430074, China}
\begin{abstract}
The quantum-to-classical transition of inflationary perturbations remains an unresolved fundamental problem, 
and quantum decoherence is one of the promising solutions. 
By considering quantum perturbations during inflation as an open quantum system interacting with its environment,
quantum decoherence can be described by the Lindblad equation. 
This formalism modifies the evolution of primordial quantum perturbations and consequently alters the power spectrum of curvature perturbations, leading to observable consequences.
In this paper, we examine the decoherence process of a polynomial attractor model
featuring an ultra-slow-roll stage, extending previous analyses limited to slow-roll scenarios.
We numerically compute the correction to the power spectrum due to quantum decoherence, 
and the results show significant modification only on large scales, with a peak generated by the decoherence correction at the minimum of the power spectrum.
Using observational constraints on the scalar spectral index and the tensor-to-scalar ratio, 
and requiring complete decoherence for relevant scales by the end of inflation, 
we obtain the constraint on the interaction parameter as $10^{-17}\text{Mpc}^{-1}<k_\gamma<0.061\text{Mpc}^{-1}$.
\end{abstract}
\maketitle

\section{Introduction}
The inflationary scenario is one of the most successful progresses in cosmology. 
It assumes that the universe experienced a rapid expansion in its very early stages, which naturally addresses the monopole, horizon and flatness problems \cite{Guth:1980zm,Linde:1981mu,Albrecht:1982wi,Starobinsky:1980te,Sato:1980yn}. 
Quantum fluctuations during inflation generate primordial density perturbations that seed the formation of large-scale structures (LSS) observed today and imprint temperature anisotropies in the cosmic microwave background (CMB) radiation
\cite{Mukhanov:1981xt,Guth:1982ec,Hawking:1982cz,Bardeen:1983qw,Mukhanov:1985rz,Sasaki:1986hm}.

However, the inflationary scenario faces several unresolved challenges, with a significant one being the quantum-to-classical transition problem \cite{Starobinsky:1986fx, Brandenberger:1990bx, Polarski:1995jg, Calzetta:1995ys,  Lesgourgues:1996jc, Burgess:1996mz, Kiefer:1998qe, Perez:2005gh, Lombardo:2005iz, Campo:2005sv, Ellis:2006fy, Martineau:2006ki, Sharman:2007gi, Kiefer:2008ku, Valentini:2008dq, Sudarsky:2009za, Asselmeyer-Maluga:2013asa, Lochan:2014dca, Rotondo:2018oln, Martin:2018zbe,  Martin:2021znx, Burgess:2022nwu, DaddiHammou:2022itk, Ning:2023ybc, Chandran:2023ogt, deKruijf:2024ufs}.
The observables such as the temperature anisotropies of the CMB and the statistical properties of the LSS, 
are inherently classical in nature. 
This raises the critical question of how the primordial quantum fluctuations that seeded these structures became classical.
The core issue stems from conventional treatments that model primordial perturbations as isolated quantum systems.
In reality, these perturbations interact with environmental degrees of freedom, leading to decoherence \cite{Zurek:1981xq, Zurek:1982ii, Joos:1984uk} that effectively converts quantum fluctuations into classical stochastic density inhomogeneities.

The dynamics of an open quantum system are described by the Lindblad master equation, which governs the time evolution of a quantum system interacting with its environment. 
Such environmental couplings induce decoherence, causing the off-diagonal elements of the system’s density matrix to diminish and driving the density matrix towards a diagonal form \cite{Burgess:2006jn, Colas:2024xjy, Burgess:2024eng}. 
This decoherence modifies primordial perturbations, leading to corrections in their power spectrum.
The decoherence timescale—over which the off-diagonal elements become negligible—depends on the interaction strength between the system and its environment: stronger coupling results in faster decoherence. 
In this paper, we impose the condition that primordial quantum perturbations at relevant scales undergo complete decoherence by the end of inflation, ensuring that these perturbations become classical thereafter.
We analyze the impact of decoherence through numerical computation of the modified power spectrum for a polynomial attractor model featuring an ultra-slow-roll stage. 
Furthermore, we derive constraints on the interaction parameter by employing the observational data P-ACT-LB-BK18 from the combination of Planck data, Atacama Cosmology Telescope (ACT) data, CMB lensing, baryon acoustic oscillation (BAO) distance measurements from Dark Energy Spectroscopic Instrument (DESI) and B-mode measurements from the BICEP/Keck telescopes (BK18) \cite{ACT:2025fju, ACT:2025tim}.

This paper is organized as follows.
In Sec. II, we present the Lindblad equation along with its exact solution under the assumption of linear interaction. 
In Sec III, we present the results of numerical calculations regarding the impact of decoherence, as well as constraints on the interaction strength. 
The paper is concluded in Sec. IV.

\section{Lindblad equation and the modified power spectrum}
The inflationary curvature perturbation is described by the gauge-invariant Mukhanov-Sasaki scalar variable $v(\eta,\bm x)$ \cite{Mukhanov:1981xt}.
The free Hamiltonian of the perturbation in Fourier space is
\begin{equation}\label{eq:mukhamiltonian}
\hat{H}_{\bm k}=
\frac{1}{2}\int_{\mathbb{R}^3}\mathrm{d}^3\bm k
\left[\omega^2(\eta,\bm k)\hat{v}_{\bm k}\hat{v}^\dagger_{\bm k}+\hat{p}_{\bm k}\hat{p}^\dagger_{\bm k}
\right],
    \end{equation}
where $\hat{v}_k$ is the Mukhanov-sasaki variable in Fourier space, $\hat{p}_k=\hat{v}_k'$ is its conjugate momentum, and a prime denotes a derivative with respect to conformal time $\eta$.
The frequency of mode $\bm k$ is given by
\begin{equation}\label{eq:omega2}
\omega^2(\eta,\bm k)=\bm k^2-\frac{(\sqrt{\epsilon_1}a)''}{\sqrt{\epsilon_1}a},
\end{equation}
where $\epsilon_1=-\dot{H}/H^2$ is the first slow-roll parameter, $H$ is the Hubble parameter and a dot indicates differentiation with respect to coordinate time $t=\int a(\eta)d\eta$.
The Hamiltonian \eqref{eq:mukhamiltonian} describes an isolated system.
However, primordial quantum perturbations are likely to interact with other degrees of freedom in the Universe. 
Therefore, a more realistic approach is to treat these perturbations as an open quantum system \cite{Martin:2018zbe,DaddiHammou:2022itk}.

For an open quantum system that interacts with its environment, 
the total Hamiltonian is
\begin{equation}
\hat{H}=\hat{H}_0+\hat{H}_{\mathrm{int}}=\hat{H}_S\otimes \hat{I}_E
+\hat{I}_S\otimes \hat{H}_E+g \hat{H}_\mathrm{int},
\end{equation}
where $\hat{H}_S$ is the Hamiltonian of the quantum system which acts in the Hilbert space $\mathcal{E}_S$, $\hat{H}_E$ is the Hamiltonian of the environment which acts in the Hilbert space $\mathcal{E}_E$, 
$\hat{H}_{\mathrm{int}}$ is the interaction Hamiltonian and $g$ is the coupling constant.
Suppose that the quantum system and its environment couple through local interactions only, 
then the interaction Hamiltonian can be written as
\begin{equation}
\hat{H}_{\mathrm{int}}=\int\mathrm{d}^3\bm x
\hat{A}\otimes\hat{B},
\end{equation}
where $\hat{A}$ acts as a local operator within the system's subsystem, while $\hat{B}$ represents a local operator associated with the environmental component.
Following Refs. \cite{Martin:2018zbe, DaddiHammou:2022itk}, we consider $\hat{A}$ to have the form
\begin{equation}
\hat{A}=\hat{v}^n,
\end{equation}
where $n$ is a constant.
By tracing out the environmental degrees of freedom from the total density matrix $\hat{\rho}_\text{tot}$, 
the reduced density matrix for the quantum system $\hat{\rho}_{S}$ becomes
\begin{equation}\label{eq:rhos}
\hat{\rho}_S=\mathrm{Tr}_E\,\hat{\rho}_\text{tot}.
\end{equation}
Using the Born and Markov approximations, the Lindblad equation 
for the reduced density matrix $\hat{\rho}_S$ is \cite{Lindblad:1975ef, Burgess:2006jn, Martin:2018zbe}
\begin{equation}\label{eq:masterx}
\frac{\mathrm{d}\hat{\rho}_S}{\mathrm{d}\eta} = i\left[\hat{\rho}_S,\hat{H}_S\right] 
-\frac{\gamma}{2}\int \mathrm{d} ^3\bm x \, \mathrm{d} ^3 {\bm y} 
\, C_{B}\left({\bm x},{\bm y}\right)
\left[\left[\hat{\rho}_S,\hat{A}(\bm y)
\right],\hat{A}(\bm x)\right],
\end{equation}
where $\gamma=2 g^2\eta_c$ with $\eta_c$ being the conformal auto-correlation time of $\hat{B}$,
and $C_B(\bm x,\bm y)=\text{Tr}_{E}[\rho_E\hat{B}(\eta,\bm x)\hat{B}(\eta,\bm y)]$ is the same-time correlation function of $\hat{B}$.
The time-dependent parameter $\gamma$ is assumed to vary with the scale factor following a power-law form \cite{Martin:2018zbe},
\begin{equation}\label{eq:gamma}
\gamma=\gamma_*\left(\frac{a}{a_*}\right)^p,
\end{equation}
where $*$ denotes a reference time when the pivot scale $k_*=0.05 \text{ Mpc}^{-1}$ crosses the horizon, and $p$ is a free parameter.
Assuming the environment to be isotropic and statistically homogeneous, $C_B(\bm x, \bm y)$ can be written as a top hat function \cite{Martin:2018zbe}
\begin{equation}\label{eq:cbdefine}
C_B(\bm x,\bm y)=\bar{C}_B\Theta
\left(\frac{a|\bm x-\bm y|}{\ell_\mathrm{E}}\right),
\end{equation}
where $\ell_{\mathrm{E}}$ is a characteristic physical correlation length and
\begin{equation}\label{eq:tophat}
\Theta(x)=
\begin{cases}
1,&\mathrm{if} \ x<1\\
0,&\mathrm{otherwise}
\end{cases}.
\end{equation}
Expanding Eq. \eqref{eq:cbdefine} in Fourier space, we obtain
\begin{equation}\label{eq:cbk}
\tilde{C}_B(\bm k)\approx\sqrt{\frac{2}{\pi}}\frac{\bar{C}_B\ell^3_\mathrm{E}}{3a^3}
\Theta\left(\frac{k\ell_\mathrm{E}}{a}\right).
\end{equation}
Considering the case where the environment consists of a heavy scalar field $\varphi$ with mass $M\gg 
H$, the interaction Hamiltonian is \cite{Martin:2018zbe}
\begin{equation}
\hat{H}_{\mathrm{int}}=\lambda\mu^{4-n-m}
\int\mathrm{d}^3\bm x\sqrt{-g}
\,\hat{\phi}^n(\eta,\bm x)
\hat{\varphi}^m(\eta,\bm x),
\end{equation}
where the parameter $\mu$ is a constant mass scale and $\hat{\phi}=\hat{v}/a$.
In this case, we have \cite{Martin:2018zbe}
\begin{align}
\bar{C}_B &= \left\lbrace \left(2m-1\right)!!-\sigma\left(m\right)\left[\left(m-1\right)!!\right]^2\right\rbrace
\left(\frac{37}{504\pi^2}\frac{H^6}{M^4}\right)^m\, ,\label{eq:barcb}\\
a\eta_\mathrm{c} & = \ell_E = 2\sqrt{2}\sqrt{\frac{\left(2m-1\right)!!-\sigma\left(m\right)\left[\left(m-1\right)!!\right]^2}{m^2\left(2m-3\right)!!}} \frac{1}{M},
\label{eq:elle}\\
\gamma &=4\sqrt{2}\sqrt{\frac{(2m-1)!!-\sigma(m)[(m-1)!!]^2}
{m^2(2m-3)!!}}\frac{\lambda^2}{M}\mu^{8-2n-2m}a^{7-2n},\label{eq:gammaexplicit}
\end{align}
where $!!$ denotes the double factorial and 
\begin{equation}
\sigma(m)=\begin{cases}
    1,\quad  \text{if } m\,\,\text{is even,}\\
    0,\quad  \text{if } m\,\,\text{is odd.}
\end{cases}
\end{equation}

For the convenience of solving the Lindblad equation \eqref{eq:masterx}, we decompose the operators $\hat{v}_{\bm k}$ and $\hat{p}_{\bm k}$ into their real and imaginary parts,
\begin{equation}\label{eq:vkpkrealimagine}
\hat{v}_{\bm k}=\frac{1}{\sqrt{2}}
\left(\hat{v}_{\bm k}^{\mathrm{R}}+i\hat{v}_{\bm k}^{\mathrm{I}}\right),\quad
\hat{p}_{\bm k}=\frac{1}{\sqrt{2}}
\left(\hat{p}_{\bm k}^{\mathrm{R}}+i\hat{p}_{\bm k}^{\mathrm{I}}\right).
\end{equation}
For a real operator $\hat{v}(\eta,\bm x)$, we have $\hat{v}_{-\bm k}=\hat{v}^\dagger_{\bm k}$, 
which implies $\hat{v}^\mathrm{R}_{-\bm k}=\hat{v}^\mathrm{R}_{\bm k}$ and $\hat{v}^\mathrm{I}_{-\bm k}=-\hat{v}^\mathrm{I}_{\bm k}$.
These relations show that not all $\hat{v}_{\bm k}$ are independent degrees of freedom, 
we need to calculate the variables $v_{\bm k}^\mathrm{R}$ and $v_{\bm k}^\mathrm{I}$ only for $\bm k\in\mathbb{R}^{3+}$.
Under the condition of linear interaction, the free Hamiltonian \eqref{eq:mukhamiltonian} can be written as 
\begin{equation}\label{eq:hamiltoniandecompose}
\hat{H}_{\bm k}^s=
\frac{1}{2}\int_{\mathbb{R}^3}\mathrm{d}^3\bm k
\left[\omega^2(\eta,\bm k)(\hat{v}_{\bm k}^s)^2+(\hat{p}_{\bm k}^s)^2
\right],
\end{equation}
where $s=\text{R, I}$.
For linear interaction $\hat{A}=\hat{v}$ ($n=1$), the density matrix can be decomposed as
\begin{equation}
    \hat{\rho}_S=
	\prod_{\bm k\in\mathbb{R}^{3+}}
	\prod_{s=\mathrm{R},\mathrm{I}}
	\hat{\rho}_{\bm k}^s.
\end{equation}
The Lindblad equation \eqref{eq:masterx} in the Fourier space for linear interaction becomes \cite{Martin:2018zbe}
\begin{equation}\label{eq:masterfourier}
\frac{\mathrm{d} \hat{\rho}_{\bm k}^s}{\mathrm{d}  \eta}=
-i\left[ \hat{\mathcal{H}}_{\bm k}^{s},\hat{\rho}_{\bm k}^{s}\right]
-\frac{\gamma}{2}(2\pi)^{3/2}
\tilde{C}_B({\bm k})\left[\hat{v}_{\bm k}^{s},\left[\hat{v}_{\bm k}^{s},
\hat{\rho}_{\bm k}^{s}\right]\right].
\end{equation}
Note that Eq. \eqref{eq:masterfourier} is valid only for linear interaction.
For the convenience of discussion, we introduce a combined interaction parameter,
\begin{equation}\label{eq:kgamma}
k_\gamma\equiv\sqrt{\frac{8\pi}{3}
\bar{C}_B{}_*\ell_E^3\frac{\gamma_*}{a_*^3}},
\end{equation}
which combines all the parameters of interaction and possesses the dimension of comoving wavenumber.

To solve the Lindblad equation, we define the eigenvectors \(\vert v_{\rm k}^s\rangle\) of the operator \(\hat{v}_{\rm k}^s\) satisfying equation \(\hat{v}_{\rm k}^s \vert v_{\mathbf{k}}^s\rangle = v_{\rm k}^s \vert v_{\rm k}^s\rangle\).
Projecting the Lindblad equation Eq. \eqref{eq:masterfourier} with the bra $\langle v_{\bm k}^{s_1}\vert$ and the ket $\vert v_{\bm k}^{s_2}\rangle$, 
we get the the solution \cite{Martin:2018zbe}
\begin{align}\label{eq:dmsolution}
& \left\langle  v_{\bm k}^{s_1}  \right\vert \hat{\rho}_{\bm k}^s\left\vert v_{\bm k}^{s_2}\right\rangle  = 
\frac{\left(2\pi\right)^{-1/2}}
{\sqrt{\left\vert v_{\bm k}\right\vert^2+\mathcal{J}_{\bm k}}}
\mathrm{exp}\left\{-\frac{
\left({v^{s_2}_{\bm k}}\right)^2+
	\left({v^{s_1}_{\bm k}}\right)^2
	+i{\left\vert v_{\bm k}\right\vert^2}^\prime
	\left[\left({v^{s_2}_{\bm k}}\right)^2-\left({v^{s_1}_{\bm k}}\right)^2\right]}
{4\left(\left\vert v_{\bm k}\right\vert^2
	+\mathcal{J}_{\bm k}\right)}\right\}
\nonumber\\ & \qquad\qquad \times
\exp\Biggl\{-\frac{\left[v^{s_2}_{\bm k}-v^{s_1}_{\bm k}\right]^2}{2\left(\left\vert v_{\bm k}\right\vert^2
	+\mathcal{J}_{\bm k}\right)}
\left(\mathcal{I}_{\bm k}\mathcal{J}_{\bm k}
-\mathcal{K}_{\bm k}^2+\left\vert v_{\bm k}^\prime \right\vert^2
\mathcal{J}_{\bm k}+ \left\vert v_{\bm k}\right\vert^2\mathcal{I}_{\bm k}
-{\left\vert v_{\bm k}\right\vert^2}^\prime \mathcal{K}_{\bm k}\right)
\nonumber\\ &  \qquad\qquad
-\frac{i\mathcal{K}_{\bm k}}{2\left(\left\vert v_{\bm k}\right\vert^2
	+\mathcal{J}_{\bm k}\right)}
\left[\left({v^{s_2}_{\bm k}}\right)^2-\left({v^{s_1}_{\bm k}}\right)^2\right]\Biggr\},
\end{align}
where $\mathcal{I}_{\bm k}$, $\mathcal{J}_{\bm k}$ and $\mathcal{K}_{\bm k}$ are defined as
\begin{align}
\mathcal{I}_{\bm k}\left(\eta\right) &\equiv
4\left(2\pi\right)^{3/2}\int_{-\infty}^\eta \mathrm{d} \eta'\gamma\left(\eta'\right) 
\tilde{C}_B\left(\bm k,\eta'\right) \mathrm{Im}^2\left[v_{\bm k}
\left(\eta'\right){v_{\bm k}^*}^\prime\left(\eta\right)\right]\, ,\\
\label{eq:defJ}
\mathcal{J}_{\bm k}\left(\eta\right) &\equiv 4\left(2\pi\right)^{3/2}
\int_{-\infty}^\eta \mathrm{d} \eta'\gamma\left(\eta'\right) \tilde{C}_B\left(\bm k,
\eta'\right)
\mathrm{Im}^2\left[v_{\bm k}\left(\eta'\right){v_{\bm k}^*}
\left(\eta\right)\right] \, ,\\
\label{eq:defK}
\mathcal{K}_{\bm k}\left(\eta\right) &\equiv 4\left(2\pi\right)^{3/2}
\int_{-\infty}^\eta \mathrm{d} \eta'\gamma\left(\eta'\right) \tilde{C}_B
\left(\bm k,\eta'\right)
\mathrm{Im}\left[v_{\bm k}\left(\eta'\right){v_{\bm k}^*}^\prime
\left(\eta\right)\right]
\mathrm{Im}\left[v_{\bm k}\left(\eta'\right){v_{\bm k}^*}
\left(\eta\right)\right]\, ,
\end{align}
and $v_{\bm k}(\eta)$ is the solution of the Mukhanov-Sasaki equation.
Using Eq. \eqref{eq:dmsolution} and the relation $\langle\hat{O}\rangle=\mathrm{Tr}(\hat{\rho}_v \hat{O})$ with $\hat{O}=\hat{v}_{\bm k}^2$, we can obtain the two-point correlation function
\begin{equation}
P_{vv}(k)=\left\vert v_{\bm k}\right\vert^2
+\mathcal{J}_{\bm k},
\end{equation}
where $\vert v_{\bm k}\vert^2$ is the standard two-point correlation and $\mathcal{J}_{\bm k}$ is the correction from the interaction with the environment. 
The modified power spectrum is
\begin{equation}
\mathcal{P}_\zeta=\frac{k^3}{2\pi^2}
\frac{P_{vv}}{2a^2\epsilon_1}=
\mathcal{P}_\zeta\vert_{\mathrm{standard}}
(1+\Delta\mathcal{P}_{\bm k}),
\end{equation}
where we take $M_{\mathrm{Pl}}^2=1$.
For linear interaction, the tensor power spectrum remains unchanged by the corrections introduced in the Lindblad equation \cite{DaddiHammou:2022itk}.
Therefore, we can utilize the spectral index and the tensor-to-scalar ratio to constrain $k_\gamma$ for linear interaction.

The interaction with the environment tends to suppress the off-diagonal elements of the density matrix when expressed in the basis of the eigenstates of the interaction operator \cite{Burgess:2006jn,Martin:2018zbe}.
From the exact solution to  Eq.
\eqref{eq:dmsolution} satisfied by the density matrix, we can study decoherence during inflation.
The parameter $\delta_{\bm k}$ defined in \cite{Martin:2018zbe} represents the reduction in the off-diagonal elements of the density matrix due to environmental interactions, and is defined as follows
\begin{equation}\label{eq:deltak}
\delta_{\bm k}(\eta) \equiv \mathcal{I}_{\bm k}\mathcal{J}_{\bm k}
-\mathcal{K}_{\bm k}^2+\left\vert v_{\bm k}^\prime 
\right\vert^2\mathcal{J}_{\bm k}+ \left\vert v_{\bm k}\right
\vert^2\mathcal{I}_{\bm k}-{\left\vert v_{\bm k}\right\vert^2}^\prime 
\mathcal{K}_{\bm k}\, .
\end{equation}
It is obvious that $\delta_{\bm k}$ is related to $k_\gamma$.
When $\delta_{\bm k}\gg1$, decoherence is expected to be fully established.
The decoherence process associated with cosmological scales of interest must have been completed by the end of the inflation, thereby establishing constraint on the interaction parameter $k_\gamma$. 
In \cite{Martin:2018zbe}, analytical expressions for the corrections to the power spectrum were derived within the slow-roll approximation.
In this paper, we numerically calculate the correction to the power spectrum for a polynomial attractor model featuring an ultra-slow-roll stage,
and compare the result with the analytical result reported in \cite{Martin:2018zbe}.

\section{Models and numerical results}
We choose the modified polynomial attractor model with the potential \cite{Wang:2024euw}
\begin{equation}\label{eq:A}
V(\psi)=V_0[1+c_4f^{-1}(\psi)+c_1f(\psi)
+c_2f^2(\psi)
+c_3f^3(\psi)]^2,
\end{equation}
where $f(\psi)=1/\psi^3$, and the parameters of the model are given in Table. \ref{tab:A1}.

From Eqs. \eqref{eq:gamma} and \eqref{eq:gammaexplicit}, we see that for linear interaction, 
\begin{equation}\label{eq:gammaprime}
\gamma(\eta')=\gamma_*\left(\frac{a(\eta')}{a_*}\right)^5.
\end{equation}
From Eqs. \eqref{eq:cbk} and \eqref{eq:barcb}, we have
\begin{equation}\label{eq:cbprime}
	\tilde{C}_B(\bm k,\eta')=
	\left(\frac{H(\eta')}{H_*}\right)^{6m}
	\left(\frac{a_*^3}{a(\eta')^3}\right)
	\sqrt{\frac{2}{\pi}}
	\frac{\bar{C}_B^*\ell_E^3}{3a_*^3}
	\Theta\left(\frac{k\ell_E}{a(\eta')}\right),
\end{equation}
where
\begin{equation}
	\bar{C}_B^*= \left\lbrace \left(2m-1\right)!!-\sigma\left(m\right)\left[\left(m-1\right)!!\right]^2\right\rbrace
	\left(\frac{37}{504\pi^2}\frac{H_*^6}{M^4}\right)^m\, .
\end{equation}
Combining Eqs. \eqref{eq:defJ}, \eqref{eq:gammaprime} and \eqref{eq:cbprime}, we have
\begin{align}\label{eq:preparameter}
	4(2\pi)^{3/2}\gamma(\eta')\tilde{C}_B(\bm k,\eta')&=
	2\left(\frac{a(\eta')}{a_*}\right)^2\left(\frac{H(\eta')}
	{H_*}\right)^{6m}\frac{8\pi}{3}
	\frac{\bar{C}_B^*\ell_E^3\gamma_*}{a_*^3}
	\Theta\left(\frac{k\ell_E}{a(\eta')}\right)
	\nonumber\\
	&=2\left(\frac{a(\eta')}{a_*}\right)^2\left(\frac{H(\eta')}
	{H_*}\right)^{6m}k_\gamma^2\,
	\Theta\left(\frac{k\ell_E}{a(\eta')}\right).
\end{align}
and
\begin{align}
	\mathrm{Im}^2\left[v_{\bm k}(\eta')
	v_{\bm k}^*(\eta)\right]=&
	\left\{\mathrm{Im}\left[\left(v^\mathrm{R}_{\bm k}(\eta')+iv^\mathrm{I}_{\bm k}(\eta')\right)
	\left(v^\mathrm{R}_{\bm k}(\eta)-iv^\mathrm{I}_{\bm k}(\eta)\right)\right]\right\}^2\nonumber\\
	=&\left(v^\mathrm{R}_{\bm k}(\eta')\right)^2
	\left(v^\mathrm{I}_{\bm k}(\eta)\right)^2+
	\left(v^\mathrm{I}_{\bm k}(\eta')\right)^2
	\left(v^\mathrm{R}_{\bm k}(\eta)\right)^2\nonumber\\
	&-2v^\mathrm{R}_{\bm k}(\eta')
	v^\mathrm{I}_{\bm k}(\eta')
	v^\mathrm{R}_{\bm k}(\eta)
	v^\mathrm{I}_{\bm k}(\eta).
\end{align}

\begin{figure}[htbp]
	\centering
	\includegraphics[width=0.7\columnwidth]{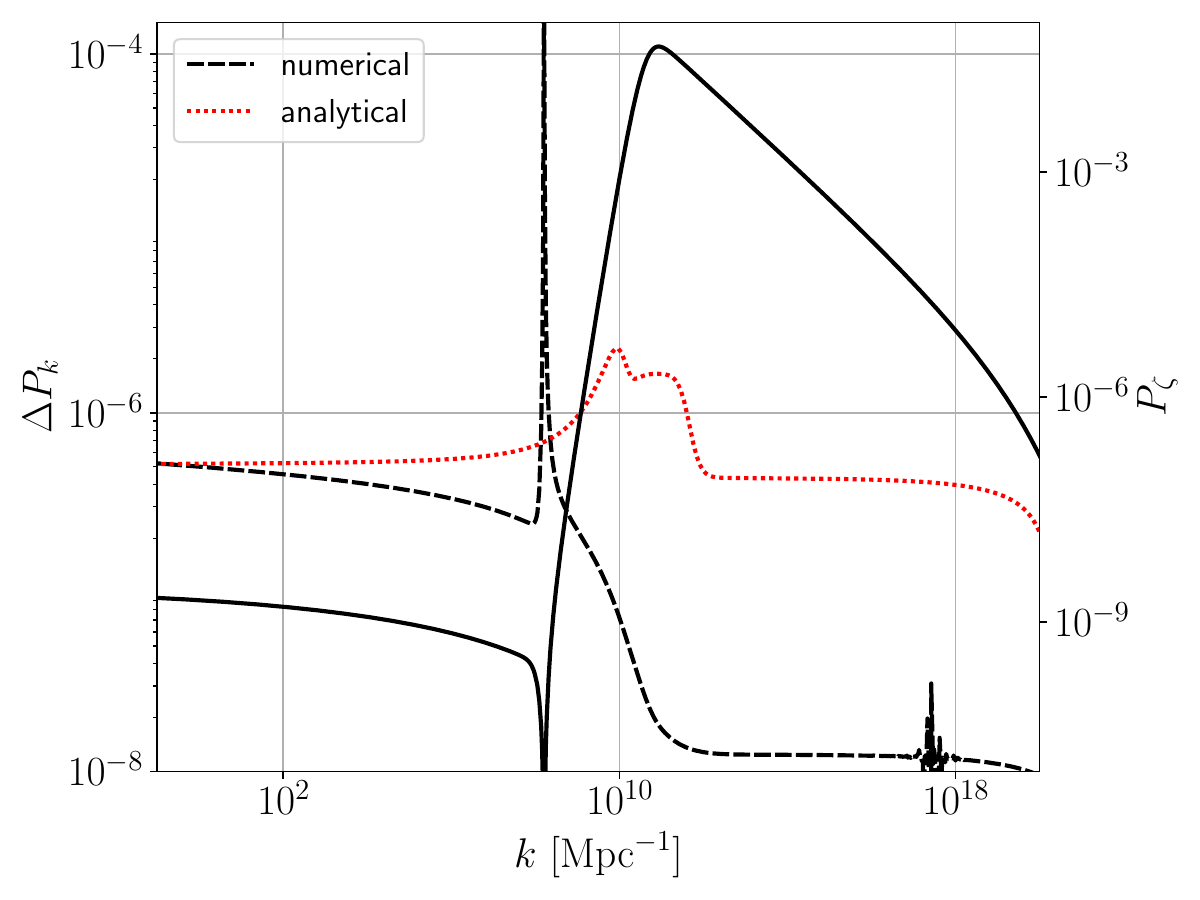}
	\caption{The power spectrum for the polynomial attractor model A. The black solid line is the modified power spectrum, the red dotted line is the correction to the power spectrum $\Delta P_k$ within slow-roll approximation, and the black dashed line is the numerical result for the correction to the power spectrum $\Delta P_k$.}
	\label{fig:dpk1order}
\end{figure}

To simplify the calculation, we take $m=3$ so that $\gamma$ defined in Eq. \eqref{eq:gammaexplicit} is independent of $\mu$.
In order to make comparison with the results in \cite{Martin:2018zbe}, we take $k_\gamma=5\times10^{-5} \text{ Mpc}^{-1}$ and $H_*\ell_E=10^{-3}$.
The top hat function in Eq. \eqref{eq:preparameter} is
\begin{equation}\label{eq:thetafunc}
	\Theta\left(\frac{k\ell_E}{a(\eta')}\right)=
	\Theta\left(\frac{k}{a(\eta')H(\eta')}
	\frac{H(\eta')}{H_*}
	H_*\ell_E\right).
\end{equation}
With these chosen parameters, we calculate the modified power spectrum numerically and the results are shown in Fig. \ref{fig:dpk1order}.

\begin{table*}
\centering
\begin{tabular}{ccccccccccc} 
\hline
Model & $V_0 $ & $\delta$ &$c_1$ &$c_4$ &  $\psi_*$  &$N$&$n_s$ &$r$ &$k_{\text{peak}}$ \\ 
\hline
A & 5.3$\times10^{-10}$ & 9.81$\times10^{-4}$  & -1.1 &2.3$\times10^{-4}$ & 4   &51.4 &0.9743 &0.017 &$8.7\times10^{10}$\\
\hline
\end{tabular}
\caption{Parameters of the model A.}
\label{tab:A1}
\end{table*}

From Fig. \ref{fig:dpk1order}, we see that at the pivot scale $k_*$, the numerical result of $\Delta P_k$ for the ultra-slow-roll model agrees with analytical one within the slow-roll approximation. 
This is because the pivot scale crosses the horizon during the slow-roll stage.
However, as the Hubble parameter gradually decreases, $\Delta P_k$ also decreases.
At the minimum of the power spectrum, $\Delta P_k$ exhibits a peak on the order of $10^{-4}$; 
yet, the correction near this peak remains negligible (approximately $\sim10^{-8}$).
This indicates that the decoherence-induced correction has an almost negligible effect on the abundance of primordial black holes and the scalar-induced secondary gravitational waves.

\begin{figure}[htbp]
	\centering
	\includegraphics[width=0.7\columnwidth]{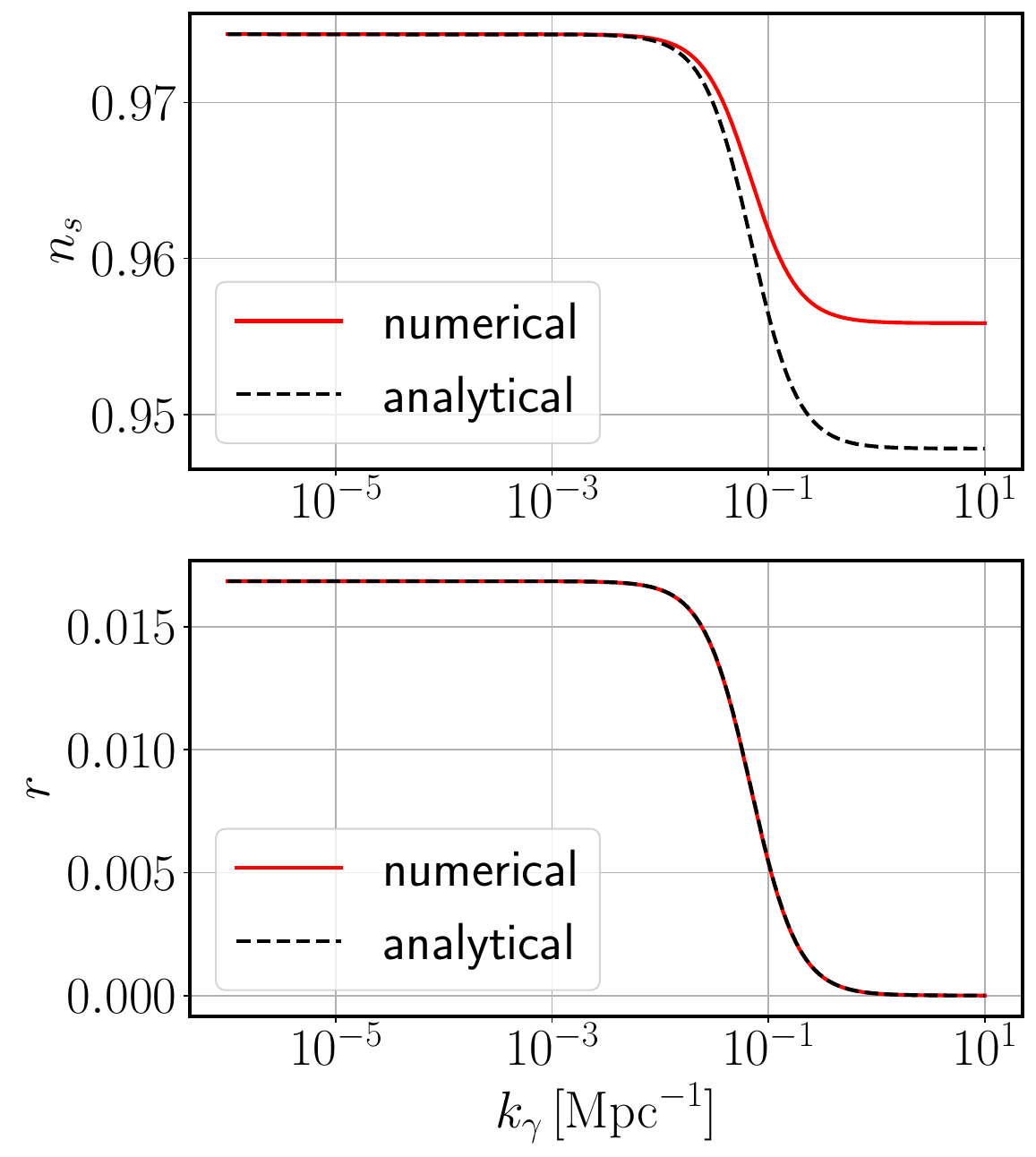}
	\caption{The behavior of the observables $n_s$ and $r$ with respect to $k_\gamma$.
    The red solid lines are the numerical results and the black dashed lines are the analytical results within slow-roll approximation.}
	\label{fig:ns_r}
\end{figure}

\begin{figure}[htbp]
	\centering
	\includegraphics[width=0.8\columnwidth]{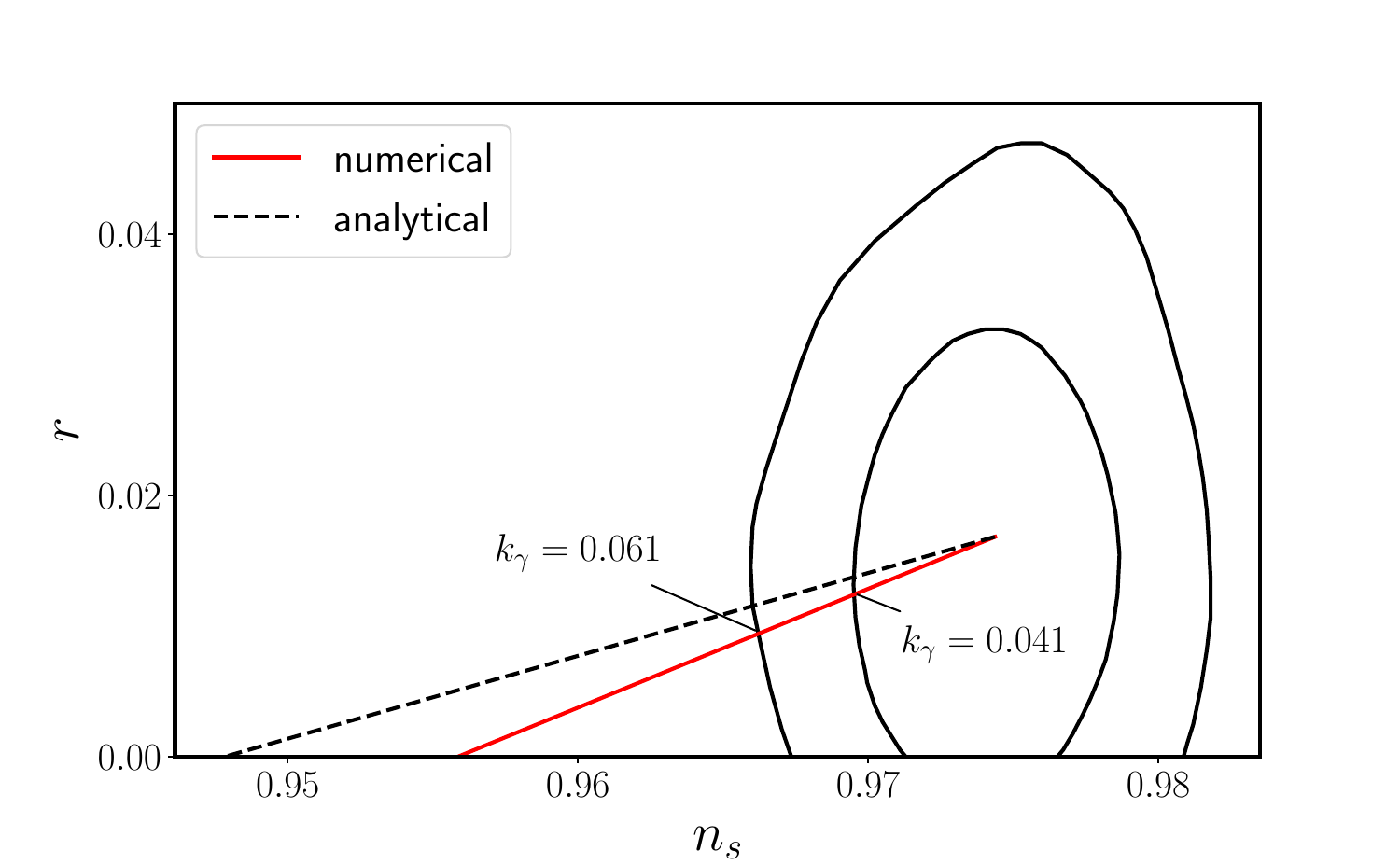}
	\caption{The spectral index and tensor-to-scalar ratio vary with $k_\gamma$.
    The red solid line represents the numerical results,
the black dashed line corresponds to the analytical results within slow-roll approximation,
and the black solid lines indicate the $1\sigma$ and $2\sigma$ constraints from P-ACT-LB-BK18. As shown by the lines,
$n_s$ and $r$ decrease as $k_\gamma$ increases.}
	\label{fig:densr}
\end{figure}

\begin{figure}[htbp]
	\centering
	\includegraphics[width=0.7\columnwidth]{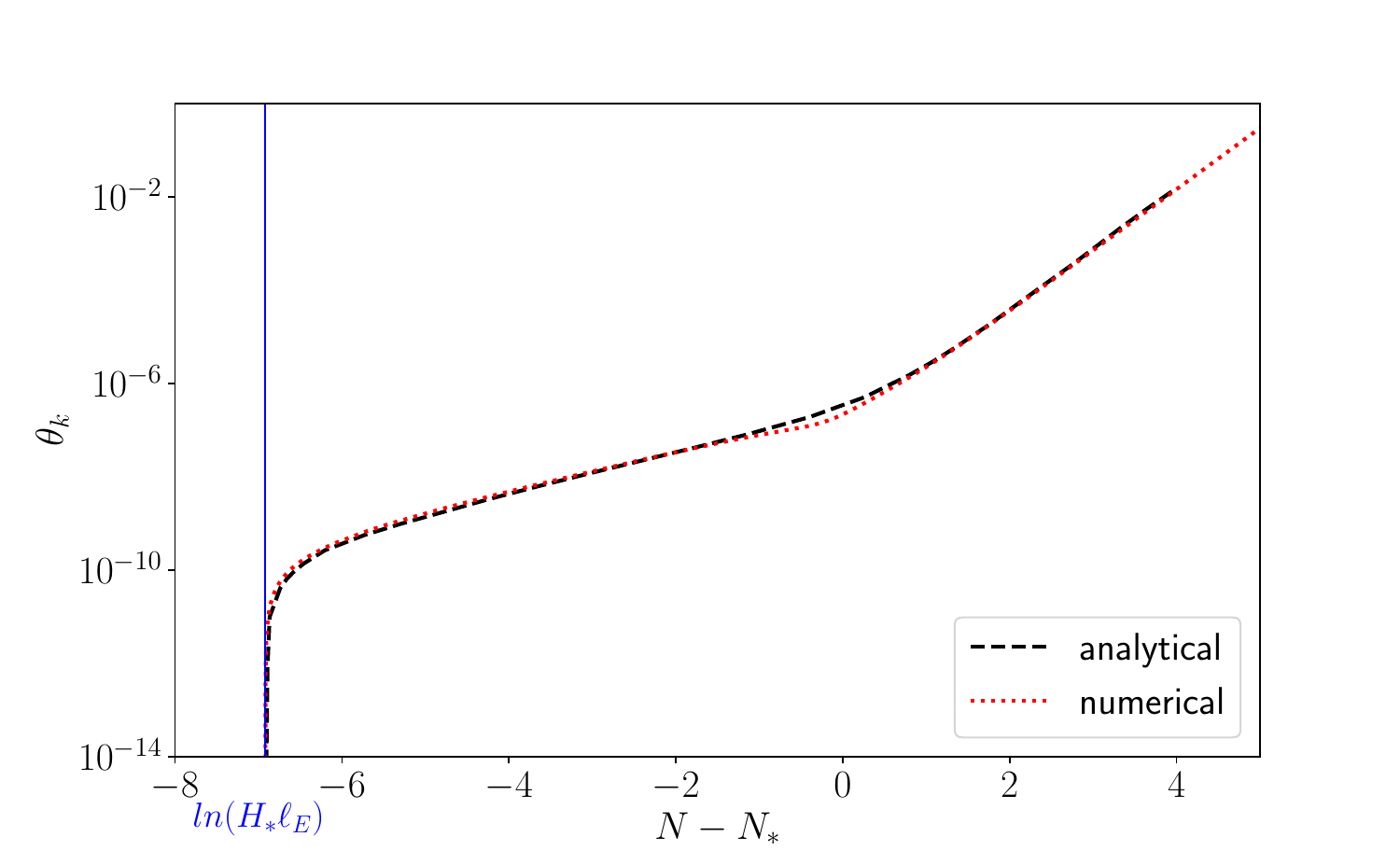}
	\caption{
    Decoherence parameter $\delta_{\bm k}$ at $k=k_*$ as a function of $N-N_*$.
    The red dotted line represents the numerical results,
    and the black dashed line corresponds to the analytical results within slow-roll approximation.
    The vertical solid line marks the e-fold value at which the evolution of $\delta_k$ begins.
	}
	\label{fig:deltak}
\end{figure}
Now we restrict $k_\gamma$ using observational constraints on the spectral tilt $n_s$ and the tensor-to-scalar ratio $r$ at the pivot scale.
The analytical results for $n_s$ and $r$ in the case of linear interaction within slow-roll approximation are \cite{Martin:2018zbe}
\begin{equation}
n_s=n_s|_{\mathrm{standard}}-\frac{\alpha}{1+\alpha}
(6m-2)\epsilon_{1*},
\end{equation}
and
\begin{equation}
r=\frac{r|_{\mathrm{standard}}}
{1+\alpha},
\end{equation}
where $\alpha=\pi k_\gamma^2/(6 k_*^2)$.
By varying $k_\gamma$,
we numerically calculate $n_s$ and $r$ at the pivotal scale and the results are shown in Fig. \ref{fig:ns_r}.
In the range $0.2\,\text{Mpc}^{-1}>k_\gamma>10^{-2}\,\text{Mpc}^{-1}$, both $n_s$ and $r$ vary significantly; outside this range, they remain nearly constant. 
For the tensor-to-scalar ratio $r$, the numerical result agrees with the analytical one within slow-roll approximation.
However, for the scalar spectral tilt $n_s$, deviations appear when $k_\gamma>10^{-2}\,\text{Mpc}^{-1}$.
In Fig. \ref{fig:densr}, we plot the results in $n_s-r$ plane along with the $1\sigma$ and $2\sigma$ constraints by P-ACT-LB-BK18.
From Fig. \ref{fig:densr}, it is evident that both $n_s$ and $r$ decrease as $k_\gamma$ increases.
The $1\sigma$ and $2\sigma$ constraints from P-ACT-LB-BK18 give upper bounds $k_\gamma<0.041\, \text{Mpc}^{-1}$ and $k_\gamma<0.061\, \text{Mpc}^{-1}$, respectively.

To constrain $k_\gamma$ using decoherence, we present the results for the decoherence parameter $\delta_{\bm k}$ at the pivot scale in Fig. \ref{fig:deltak}.
Since the pivotal scale crosses the horizon during the slow-roll phase, the numerical results agree well with the analytical approximations.
When $N-N_*\geq 4$,  where $N$ is the number of e-folds before the end of inflation and $N_*$ is the number of e-folds remaining at the horizon crossing of the pivot scale, $\delta_{\bm k}$ increases exponentially.
We require decoherence to be complete for both the scales $k_*$ and $k_{\mathrm{peak}}$ by the end of inflation,
leading to the lower bound $k_\gamma \gg 10^{-17}\, \text{Mpc}^{-1}$.
Compared to the results obtained in Ref. \cite{Martin:2018zbe}, 
our bound is tighter because we demand that perturbations not only at CMB scales but also at the peak of the power spectrum fully decohere by the end of inflation.

\section{Conclusion}
In this paper, we examine the effects of decoherence on the power spectrum in an inflationary model featuring an ultra-slow-roll phase. Decoherence, arising from interactions between the quantum system and its environment, is described by the Lindblad equation. Previous studies, employing the slow-roll approximation, calculated the interaction’s impact on the power spectrum and placed preliminary constraints on both the interaction parameter $k_\gamma$ and decoherence parameter $\delta_{\bm k}$.

Using solutions to the Lindblad equation for linear interactions from Ref. \cite{Martin:2018zbe}, 
we numerically compute the corrections to the power spectrum of a polynomial attractor model and constrain the interaction parameter $k_\gamma$ using observational data.
We find that decoherence significantly modifies the power spectrum primarily on large scales, consistent with analytical results obtained under the slow-roll approximation. 
As the scale decreases, the corrections become smaller, but a peak appears at the trough of the power spectrum.

By analyzing the impact of $k_\gamma$ on observables such as the scalar spectral tilt $n_s$ and the tensor-to-scalar ratio $r$, we establish an upper limit of $k_\gamma<0.061$ \,Mpc$^{-1}$ from the P-ACT-LB-BK18 data.
We also calculate the decoherence parameter $\delta_k$ at the pivot scale, which is used to characterize the degree of completion of decoherence. 
Requiring complete decoherence at both the pivot scale $k_*$ and the peak scale $k_{\mathrm{peak}}$ by the end of inflation
leads to a tighter lower bound, $k_\gamma \gg 10^{-17}\, \text{Mpc}^{-1}$.

In conclusion, we constrain the interaction parameter within the range $10^{-17}\text{Mpc}^{-1}<k_\gamma<0.061\text{Mpc}^{-1}$.

\begin{acknowledgments}
This work is supported in part by the National Key Research and Development Program of China under Grant No. 2020YFC2201504.
\end{acknowledgments}

%

\end{document}